\documentclass[]{spie}  

\usepackage{amsmath,amsfonts,amssymb}
\usepackage{graphicx}
\usepackage[colorlinks=true, allcolors=blue]{hyperref}
\usepackage{float}
\usepackage{ifthen}

\newboolean{eac4}
\setboolean{eac4}{true}

\title{Connecting Polarization to Exoplanet Yield Calculations for HWO}

\author[a]{Jaren N. Ashcraft    \footnote{NASA Hubble Fellow} }
\author[b]{Scott D. Will}
\author[a]{Maxwell A. Millar-Blanchaer}
\author[b]{Breann Sitarski}
\author[b]{Chris Stark}

\author[c]{Ewan S. Douglas}
\author[d]{John Krist}
\author[d]{Brian Kern}
\author[b]{Manuel Quijada}

\affil[a]{Department of Physics, University of California, Santa Barbara, CA, 93106, USA}
\affil[b]{NASA Goddard Space Flight Center, Greenbelt, MD, 20771}
\affil[c]{Department of Astronomy and Steward Observatory, University of Arizona, 933 N. Cherry Ave., Tucson, AZ 85721, USA}
\affil[d]{NASA Jet Propulsion Laboratory, California Institute of Technology, 4800 Oak Grove Drive, Pasadena, CA 91109, USA}
\authorinfo{Further author information: (Send correspondence to J.N.A.)\\J.N.A.: E-mail: jarenashcraft@ucsb.edu} 

\begin{document} 
\maketitle
\begin{abstract}
The Habitable Worlds Observatory (HWO) aims to enable the detection and characterization of Earth-like planets around Sun-like stars to search for possible signs of life elsewhere in our universe. This requires an incredibly sensitive coronagraph instrument that suppresses the light from the star by a factor of 10 billion, which must contend with error terms that have not previously limited high-contrast instrumentation at lower levels of starlight suppression. Polarization aberrations are one such source of error that is particularly problematic for coronagraphy on a large space telescope. Optical rays in large, compact astronomical observatories can have large changes in angle of incidence over the beam, which induce polarization aberrations that decrease sensitivities to faint signals at small angular separations. Limiting variation in angles of incidence along the optical path could lead to longer, less stable observatories. This could negatively impact the total number of exo-Earths HWO would be able to detect. This study links open-source physical optics modeling tools to an exoplanet yield optimizer to understand how polarization aberrations influence science return for HWO. We also explore how polarization aberrations scale with change in angle of incidence, which could drive the primary-secondary mirror distance and overall observatory stability. In the visible, we find that decreasing the EAC-1 barrel from 16m to 12m results in $\approx 10^{-10}$ contrast at the IWA where we expect exo-Earths to be. In the UV we appear to be less sensitive to polarization because exo-Earths are farther from the IWA. We also find a limited range over which the design reference mission of EAC-1 can be optimized to compensate for polarization aberrations using altruistic yield optimization. We then report on mitigation strategies to minimize the presence of polarization aberrations in HWO.
\end{abstract}


\section{INTRODUCTION}

Directly imaging and characterizing Earth-like planets around Sun-like stars requires unprecedented starlight suppression to achieve $10^{-10}$ contrast levels.
Polarization-dependent wavefront errors, also called polarization aberrations, can be a limiting factor in achieving these deep contrasts, and can drive the size of space telescopes.
These wavefront errors arise from any reflection at nonzero angles of incidence and transmission through birefringent materials.
Thus, varying angles of incidence across the beam, coating variations between optics, and different coating recipes (i.e., aluminum with dielectric overcoats and silver with overcoats), yield differential complex amplitude between polarization states, as well as cross-talk terms that vary spatially\cite{Breckinridge_2015}.
The spatial variations project primarily into tip-tilt and astigmatism modes.
As a result, the contrast degradations they produce are concentrated near the coronagraph’s inner working angle, where most planets are expected to be found at visible wavelengths\cite{anche_2023, ashcraft_gsmts_2025, Stark_14}.
This limits the sensitivity of the telescope to possible Exo-Earths around solar-type stars.
Polarization aberrations are not fully correctable, and require polarizing filters \cite{Breckinridge_2015, krist_numerical_2023} to simplify the wavefront control problem.
This strategy, called polarization splitting, mitigate polarization aberrations by sending orthogonally polarized light to different coronagraphs. 
This permits independent WFSC systems to correct for orthogonal polarization states simultaneously.
While this eases the wavefront control of polarization aberrations\cite{ashcraft_6mst_2025}, it comes at the cost of halving the coronagraph throughput and doubling the payload size of the coronagraph, which adds risk and expense to a spaceborne mission.

However, with new coronagraph masks and coating recipes, especially aluminum-based coatings that are enhanced for the far ultraviolet wavelength range, those constraints need to be reassessed with respect to the architectures for the Habitable Worlds Observatory (HWO).
The HWO project office developed three point designs, called exploratory analytic cases (EACs), to perform exploratory trades.
Understanding the contribution of polarization as designs for HWO mature is crucial so that results from this study can be incorporated into future architecture studies and the design of HWO\cite{Feinberg2024}.
This study explores the relationship between the size of the optical telescope assembly (OTA) and the resulting contrast limits imposed by polarization aberration.

We report on the findings of a parameter study to determine the sensitivity of HWO coronagraphs to polarization aberrations.
In Section \ref{sec:methods} we report on the integration of physical optics models with the Framework for Remote Implementation of Demand-based Altruistic Yields (FRIDAY), an interface to Stark et al.'s Altruistic Yield Optimizer (AYO)\cite{Stark_14}. 
FRIDAY was created by Stark to grant members of the community remote access to a server that runs AYO. 
The user supplies the server with a \texttt{.ayo} job script and a large file containing multiple realizations of the coronagraph's PSF as a function of field position.
Using these data, FRIDAY can remotely execute AYO and return yield analysis plots to the user.
This modeling pipeline enables us to directly connect the presence of polarization aberrations to Earth-like exoplanet yields for a HWO-like mission.
In Section \ref{sec:results_eac1} we report on efforts in polarization aberration modeling for the EAC-1 architecture for wavelengths spanning the UV to the IR, finding that polarization aberrations are most limiting in visible wavelengths. In Section \ref{sec:results-eac4} we extend these analyses to the EAC-4 design, where we explored 3 point designs that used "maximum AOI" as a design specification. 
In Section \ref{sec:conclusions} we discuss the implications of this work for future iterations of NASA's HWO, and methods to mitigate the impact of polarization aberrations.

\section{METHODS}
\label{sec:methods}

\subsection{Integrating Optical Aberrations into Science Yields}
To model the EAC polarization aberrations we use \verb|Poke| \cite{Ashcraft_poke_2023}, an open-source ray-based physical optics package.
\verb|Poke| is written in Python, and uses the CODE V and Zemax OpticStudio API's to produce ray data for a given optical system.
\verb|Poke| then carries out the polarization ray tracing calculus described in Chipman, Lam, and Young \cite{CLY} to compute the polarization aberrations expressed in the local coordinates of a given instrument.
In this way, \verb|Poke| open-sources the polarization aberration computation while being compatible with ray-tracing models that are industry-standard.
\verb|Poke| has been used in prior studies in support of HWO \cite{ashcraft_6mst_2025} and the ground-based 30m extremely large telescopes \cite{anche_2023, ashcraft_gsmts_2025}. For a brief overview on the polarization aberration computation, refer to the introduction of Ashcraft et al. 2025 \cite{ashcraft_6mst_2025}. A more exhaustive review can be found in the work of Yun and Chipman\cite{Yun:11_1, Yun:11_2}. 

To model the influence of diffraction effects in the OTA and the coronagraph, we use the \verb|Jax|-based scalar diffraction tool \verb|hwo_sim|, which implements the algorithm presented in Will and Fienup 2019\cite{Will_polarization_luvoir}.
The scalar diffraction simulation begins by propagating wavefront error through a plane-to-plane model of the telescope optical system.
The wavefront error maps from each optic are generated by a power spectral density specified by the EAC-1 architecture team.
The total phase and amplitude aberrations are then back-propagated to the coronagraph entrance pupil. 
Here the aberrated wavefront is conjugate to an in-pupil $96 \times 96$ model of a Deformable Mirror (DM) with an 85mm clear aperture.
After applying the influence of this DM, the resulting wavefront is free-space propagated to an identical out-of-pupil DM located 1 meter in front of the in-pupil DM that enables the simultaneous correction of phase and amplitude \cite{Pueyo:09}.
The corrected wavefront is then propagated through the coronagraph (described in more detail in Section \ref{sec:optical_params_eac1}) and to the science focal plane. 
Using this forward model, we employ Jacobian-free wavefront control as described in Will and Fienup 2021 \cite{Will_jacobian_2021} to minimize the energy in a dark hole region from $3 \rightarrow  21\lambda_0 / D$. 

Both \texttt{poke} and \verb|hwo_sim| have been validated against methods in the literature and existing software tools. \texttt{Poke}’s automated unit testing includes tests against examples given in the Polarized Light and Optical Systems textbook by Chipman, Lam, and Young \cite{CLY}. \verb|hwo_sim| has been validated to exo-earth contrast levels against PROPER, which has in turn been validated with the Roman Coronagraph in vacuum at NASA JPL. 

The final component of the modeling ensemble is to connect the integrated optical models to a science metric.
Exoplanet yield is one such metric of interest to the HWO community for determining the ability of a mission to detect Earth-like exoplanets.
Exoplanet yield calculators are tools that simulate the design reference mission (DRM) of an observatory's instrument campaign\cite{Stark_14}.
EXOSIMS\cite{exosims} and the Altruistic Yield Optimizer (AYO)\cite{Stark_14} are two of the most well-known of these calculators.
AYO was constructed following the work of Stark et al. 2014\cite{Stark_14} which indicated that future spaceborne direct imaging missions require optimization of the DRM to maximize exoplanet yield.
This includes parameters such as exposure time, number of re-visits, time between visits, and the specific stars selected for observation.
In the exoplanet yield modeling space, exo-Earth yield ($Y$) is defined by the expression in Equation \ref{eq:yield_def},

\begin{equation}
    \label{eq:yield_def}
    Y = \eta_{\circ} \circ \sum_{i}^{N_{sys}} C_{i}.
\end{equation}

Where $\eta_{\circ}$ is the estimated fraction of $N_{sys}$ extrasolar systems which have Earth-like planets in the habitable zone, and $C$ is the Habitable Zone (HZ) completeness of the system, which is the probability of detecting the exoplanet given that the exoplanet \emph{is in the system}.
AYO deposits a planet in the HZ of each simulated system, and performs a simulated observation of each system in the target list.
By optimizing the detection and characterization time, and number of observations of a given extrasolar system, AYO maximizes $Y$.
This method permits AYO to compensate for modest amounts of optical aberration in the coronagraph by adjusting the parameters of the DRM.
This leads us to ask the question, \emph{how sensitive is yield to polarization aberration}?

\begin{figure}[ht]
    \centering
    \includegraphics[width=\linewidth]{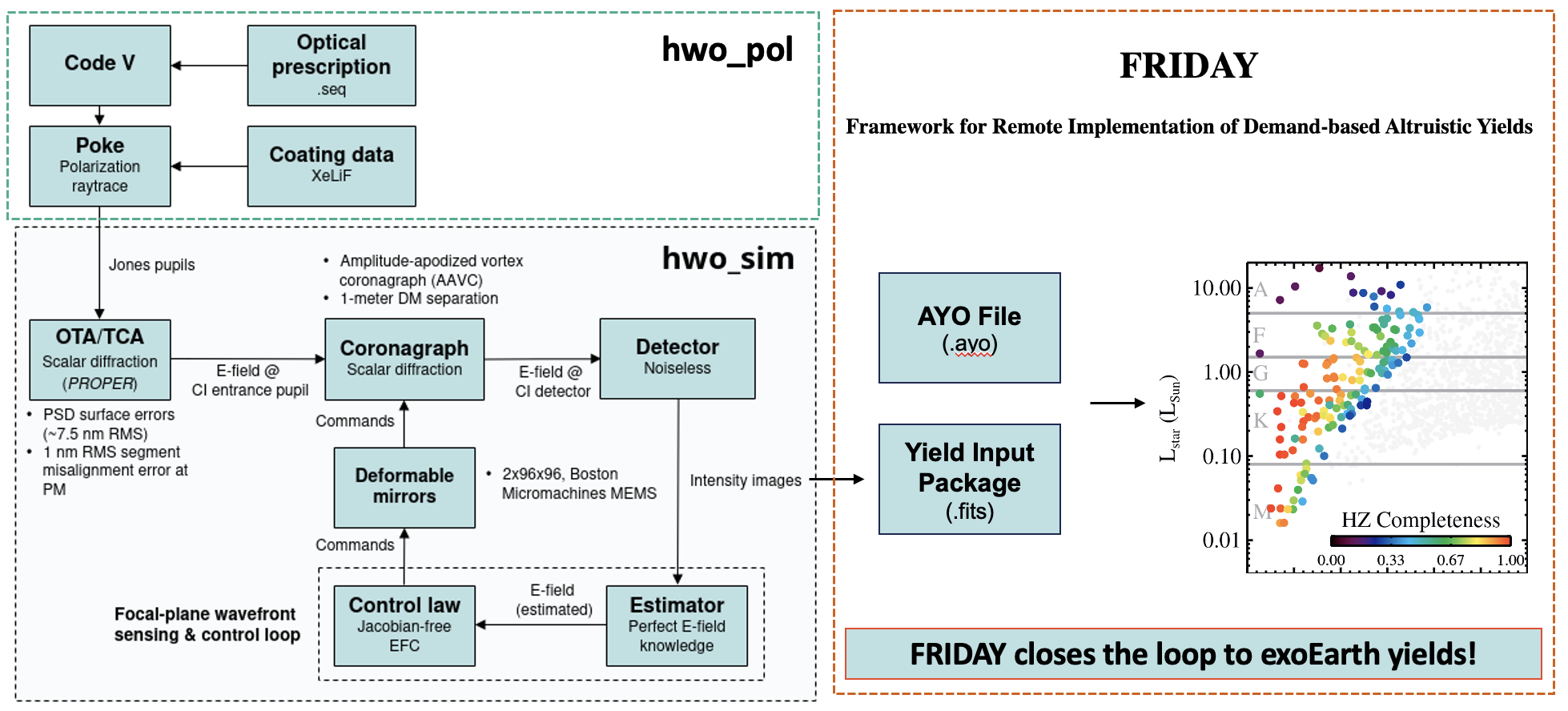}
    \caption{Overview of the simulation pipeline components used in this work. The modeling begins with \texttt{hwo\_pol}, a \texttt{Poke} front-end we constructed tailored to modeling the EACs. The ray trace file and coating data are fed to \texttt{Poke} which computes the Jones pupils and saves them as a FITS file. These data are loaded into \texttt{hwo\_sim} and applied on top of the phase and amplitude errors encountered by propagating through the OTA and tertiary collimator assembly (TCA, mirrors 3-5). The now polarization aberrated field is propagated through a scalar diffraction model, subject to jacobian-free wavefront control with perfect knowledge of the electric field. Using this control simulation, we can save a series of intensity images into a Yield Input Package, which is sent to the FRIDAY AYO calculator to close the loop to exoEarth yields.}
    \label{fig:simulation_pipeline}
\end{figure}

Our first goal in this study was to provide a straightforward modeling pipeline to connect the three codes to answer this question.
In this study, the ray-trace files are generated with CODE V by the HWO optical design team and saved as sequence (\verb|.seq|) files. 
\verb|Poke| generates the Jones pupils given the ray-trace file and exports them in the FITS file format.
These are loaded into \verb|hwo_sim| and propagated through the OTA and coronagraph to arrive at a polychromatic electric field estimate at the focal plane.
We use \verb|hwo_sim| to generate the 556 PSF realizations per wavelength for AYO's yield input packages (YIPs) for the ultimate computation of exoplanet yield.
AYO will then operate using the YIPs to determine our exoplanet yield metric.
Note that it is possible to generate YIPs via the Coronagraph Design Survey's pipeline \cite{cds_pipeline}, which contains a standardized set of analysis tools for evaluating coronagraph robustness.
A review of the simulation pipeline constructed for this work is shown in Figure \ref{fig:simulation_pipeline}.

To minimize the computational complexity of simulating high-order wavefront sensing and control (HOWFSC) for every forward model, we derive analytically what a pairwise-probing type estimator would see for a field subject to polarization aberrations, and apply the correction to the polarization aberrations. This allows us to perform one HOWFSC simulation for a scalar case, and apply the polarization aberrations to it for rapid exploration of coronagraph sensitivity to polarization. This derivation is included in a companion paper by Will et al.\cite{Will_2025_idealized}, but the results are summarized here for completeness. In the presence of polarization aberrations, the intensity at the coronagraphic focal plane is proportional to a sum of the propagated elements of the Jones pupil. The Jacobian computed for EFC back-propagates the observed intensity to an estimate of the electric field in the coronagraph's exit pupil. This linear operation transforms intensities to electric fields, which in the presence of polarization aberrations, transforms a sum of intensities on the focal plane to a sum of electric fields in the exit pupil. Consequently, a pairwise-type estimator will see polarization aberrations as a coherent sum of the Jones pupil elements in the exit pupil.

To compute the ideal focal-plane electric field seen by a pairwise-type estimator,
\begin{enumerate}
    \item Calculate the piston phase component of each Jones pupil element: $exp(i\phi_m)$
    \item Calculate the scaling factor for each Jones pupil element, accounting for throughput $T$: $c_m = \frac{rms(|J_m|)}{\sqrt{T}}$
    \item Propagate each piston-subtracted element through the coronagraph: $E_m = \mathcal{C}[J_m / exp(i\phi_m)]$
    \item Calculate the coherent sum: $\hat{E} = \frac{1}{2}\Sigma_m c_m E_m$
\end{enumerate}

The final term $\hat{E}$ is what an ideal pair-wise estimator would observe from a polarization aberrated field in the pupil plane, which we apply the correction for on the deformable mirror.
We apply this correction after digging the dark hole for the scalar HOWFSC
Armed with an expeditious integrated modeling pipeline that encapsulates polarization, diffraction, and their impact on exo-Earth yields, we are able to conduct rapid trade studies in support of HWO.

\subsection{Optical Parameters for EAC-1}
\label{sec:optical_params_eac1}
\begin{figure}[ht]
    \includegraphics[width=\textwidth, trim={3cm 0cm 3cm 0cm}]{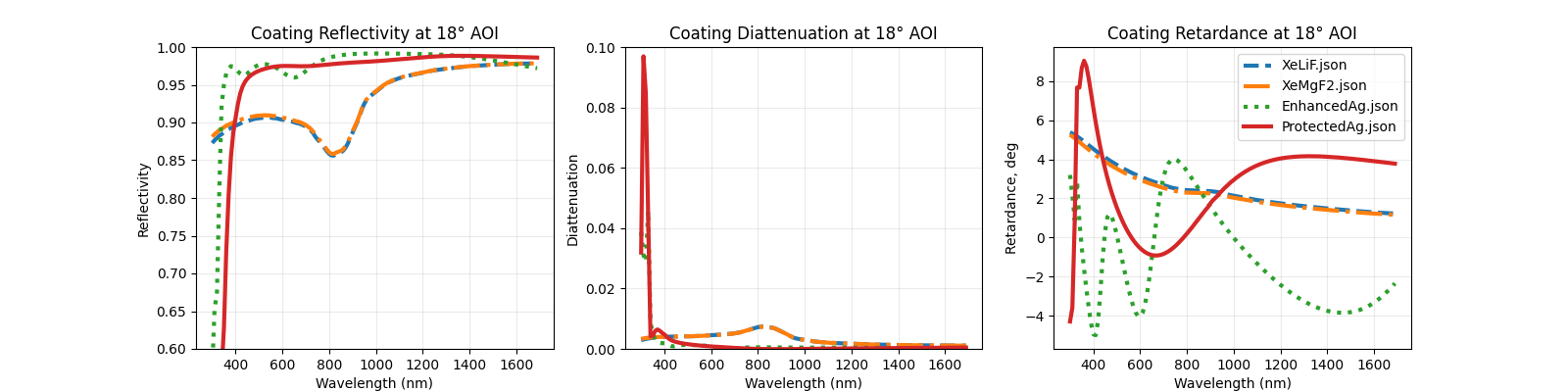}
    \caption{The reflectivity (left) diattenuation (center) and retardance (right) of the coatings considered in this study. The aluminum-based XeLiF (blue) and XeMgF2 (orange) exhibit similar spectral characteristics across the all three plots. The Enhanced (green) and Protected (red) Silver plots, in contrast, exhibit higher-order spectral characteristics. These data assume an angle of incidence of $18^{\circ}$.}
    \label{fig:coating_nominal}
\end{figure}

The EAC-1 optical telescope assembly (OTA) is an off-axis three-mirror-anastigmat (TMA) design with a 6 meter (inscribed\cite{feinberg2026habitableworldsobservatorysconcept}) primary mirror diameter, where the tertiary mirror feeds the coronagraph. 
The primary-secondary mirror separation for the nominal design is 16 meters, and is a parameter that will be varied to explore how it drives polarization aberrations.
The exit pupil of the TMA is relayed to a fast-steering mirror which forms the entrance pupil of the coronagraph.
The baseline coronagraph for these analysis is the EAC-1 coronagraph, a charge-6 amplitude-apodized vortex coronagraph (henceforth, AVC).
This design was developed by Susan Redmond and Dimitri Mawet for EAC-1, and is assumed to be achromatic and polarization-independent.
While technologies are in development that aim to construct such a coronagraph\cite{Doelman_2022, Palatnick:24}, this is not yet a physically realizable mask.
This model provides an upper bound on how we can expect vortex coronagraphs to perform subject to polarization aberrations. 
One critical modification we assume in this analysis that departs from the EAC-1 design is that \emph{we do not employ polarization splitting}.
We assume that light from all polarization states hits the coronagraph and final science focal plane.
The polarization splitting architecture was suggested to mitigate polarization aberration, but requires the heavy penalty of sacrificing half of the throughput and limiting the accuracy of polarimetric science.
The results of this analysis should then inform if polarization splitting is necessary for an EAC-1-like observatory architecture.
To compute the polarization aberrations from an EAC-1-like observatory, we must simulate the coatings for every mirror in the optical train.
Novel coatings are undergoing development to enhance HWO's capabilities in the ultraviolet (UV), and their contributions to polarization aberrations must be understood.
In this study we compare a xenon-fluorinated lithium fluoride (XeLiF) and xenon-fluorinated Magnesium fluoride (XeMgF2) dielectric layer over aluminum the coating for the primary and secondary mirror to understand how similar coatings may behave differently. The refractive indices for the XeLiF and XeMgF2 coatings were obtained from measurements conducted at the coating facility at NASA Goddard Space Flight Center\cite{Quijada_2025}. The refractive indices for the protected and enhanced silver coatings were provided via a private communication with a vendor. The reflectivity, and nominal polarization aberrations (diattenuation, retardance) are shown in Figure \ref{fig:coating_nominal}.

The aluminum-based coatings are intended to be deposited on the Primary and Secondary mirrors of HWO to enable science in the ultraviolet. For the remaining optics in the coronagraph train, we consider two different silver-based coatings.
The enhanced silver coating in Figure \ref{fig:coating_nominal} was designed for a coronagraph in the ultraviolet, utilizing more dielectric layers to enhance UV throughput.
The protected silver coating in Figure \ref{fig:coating_nominal} is a more general-purpose coating for coronagraph optics in the visible and infrared given its higher reflectivity in those wavelengths. The simulations for the ultraviolet in this study use the enhanced silver coating for all of the optics after the secondary mirror. The remaining wavebands considered in this study spanning the visible to the near-infrared use the more general-purpose protected silver coating.

With the assumptions stated in this Section, it is critical for the reader to consider the degree of realism in this study. We aim to isolate the contribution of polarization aberrations to normalized intensity degradation for an HWO-like coronagraph, and by doing so make the following unrealistic assumptions:

\begin{enumerate}
    \item A perfectly achromatic, and scalar vortex coronagraph
    \item Noiseless wavefront sensing
    \item Ideal wavefront control of polarization aberrations
\end{enumerate}

The results presented in Sections \ref{sec:results_eac1}-\ref{sec:results-eac4} should be considered a lower bound to the performance achieveable subject to polarization aberrations, as we don't consider the conflation of these errors with other effects that degrade the coronagraph's normalized intensity. Rather, this study aims to rapidly explore the space of polarization aberrated normalized intensity so that we can later revisit specific cases of interest with a greater degree of realism.

\section{RESULTS - EAC-1}
\label{sec:results_eac1}
\begin{figure}[ht]
    \includegraphics[width=\textwidth]{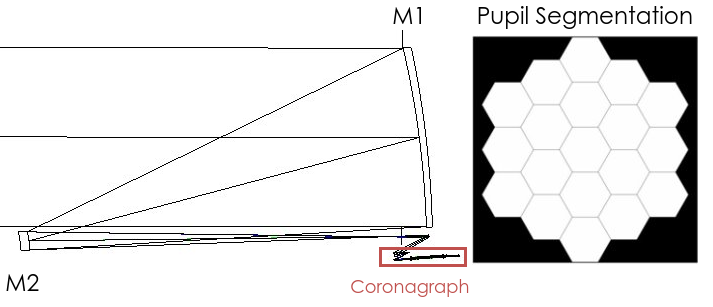}
    \caption{(Left) Ray diagram of the EAC-1 observatory and coronagraph which is packaged under the primary mirror (M1). The pupil segmentation is hexagonal (Right), with a 7200mm outer diameter, that gets circularized at the coronagraph entrance pupil to minimize the diffraction from the hexagonal array.}
    \label{fig:EAC1_raytrace}
\end{figure}

A trade of particular interest to the architecture of HWO is how the length of the telescope barrel translates into polarization aberrated normalized intensity.
To conduct this exploration, we consider Exploratory Analytical Case 1 (EAC-1). This observatory, shown in Figure \ref{fig:EAC1_raytrace}, utilizes an off-axis design to minimize the diffraction from secondary mirror obscurations. We compute the polarization aberrations of M1 through the optic immediately before the coronagraph's focal plane mask using \verb|Poke|, and then back-propagate the results to the coronagraph entrance pupil for propagation using \verb|hwo_sim|. For a more detailed description of the coronagraph layout, please consult Noecker et al. (this issue)\cite{Noecker_hwo}.

\begin{table}[h]
    \centering
    \begin{tabular}{c|cccccccccc}
    \hline
        Barrel & 16 & 15 & 14 & 13 & 12 & 11 & 10 & 9 & 8 & 7  \\
    \hline
        Maximum AOI [deg] & 13.0 & 13.9 & 14.8 & 15.9 & 17.2 & 18.6 & 20.4 & 22.5 & 25.0 & 28.0 \\
        M1 Focal Ratio & 2.96 & 2.78 & 2.59 & 2.41 & 2.22 & 2.04 & 1.85 & 1.67 & 1.48 & 1.30 \\
        $\Delta$ AOI [deg] & 11.4 & 12.1 & 12.9 & 13.8 & 14.9 & 16.1 & 17.7 & 19.4 & 21.5 & 24.0 \\
    \hline 
    \end{tabular}
    \caption{Parameters relevant to the investigations in this study across the 10 designs analyzed for EAC-1. M1 focal ratio and $\Delta$ AOI are directly responsible for the anticipated polarization aberrations, while maximum AOI was a parameter of interest to the optical design team.}
    \label{tab:aoi_params}
\end{table}

A smaller barrel leads to lower size, weight, and power (SWAP), drastically reducing the cost of the mission.
The cantilevered design of the secondary mirror means that the requirements on stability are larger the further it is separated from the primary mirror.
However, bringing the mirrors closer together requires ``faster" (i.e. lower $F/\#$) optics, which directly correlate to increased polarization aberrations given the increased variation in angle of incidence over the primary and secondary mirror optcs.

\begin{figure}[ht]
    \centering
    \includegraphics[width=\linewidth, trim={0cm 0cm 0cm 0cm}]{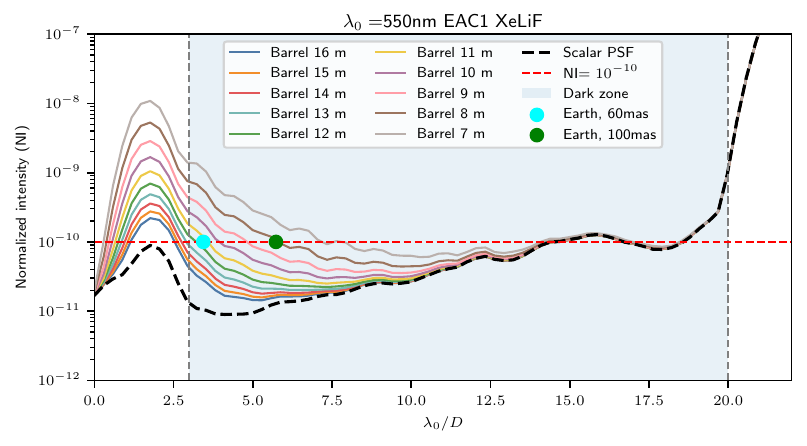}
    \caption{Azimuthally-averaged PSF profiles for each of the 10 optical designs used in this study, with polarization aberrations evaluated in the visible for a $20\%$ bandpass centered on 550nm. The profiles are plotted as a function of the center wavelength divided by the entrance pupil diameter, $\lambda_{0} / D$. To keep track of the chromatic scaling of the PSF, we also drop two fictitious ``Earths" at 60 and 100 mili-arcseconds (mas) to qualitatively assess how such exoplanets might be limited by polarization aberrations. In these data, we observe the 60mas Earth is right after the IWA of the coronagraph. Consequently, it is approximately the same normalized intensity as the degraded focal plane residuals from the 12m barrel polarization aberrations. The 100mas planet near 5 $\lambda_0 / D$ is nearly the same normalized intensity as the 8m barrel polarization aberrations. These data are in dual-polarization, meaning that no polarization splitting was done to minimize the presence of polarization aberrations. These data suggest that EAC-1 could in principle tolerate moving to a shorter barrel, but future work containing a more realistic model of the coronagraph and HOWFSC is required to make this assertion.}
    \label{fig:550nm_focalplane_eac1}
\end{figure}

To constrain this design parameter we construct 10 modifications to the EAC-1 designs with progressively shorter barrels in 1 meter increments, spanning 16m to 7 meters. The parameters relevant to polarization aberrations are given in Table \ref{tab:aoi_params}.
The radii of curvature and conic constants of the primary and secondary were set to variable, and we re-optimized the design to reduce the separation of the two mirrors in 1 meter increments.
The optimization was constrained by maintaining the focal ratio of the beam after the secondary, and that the static wavefront error did not change substantially.
We consider the scalar wavefront error to be corrected either in alignment, or with the onboard adaptive optics, so they are not considered in this study so as to not conflate them with polarized aberrations.
For each of the 10 designs, we apply the idealized correction described in Will et al.\cite{Will_2025_idealized} (and in brief in Section \ref{sec:methods} of this work) and simulate the coronagraphic residuals after HOWFSC.
The azimuthal averages of these image simulations are shown in Figure \ref{fig:550nm_focalplane_eac1}. 
These curves are normalized to the maximum of an on-axis PSF with the focal plane mask removed.
We observe a trend that is consistent with prior investigations into polarization aberration\cite{ashcraft_6mst_2025, anche_2023, ashcraft_gsmts_2025, Will_polarization_luvoir, Breckinridge_2015}, the prominent errors arise near the inner working angle, and get worse as optics get faster (i.e. shorter barrel length).
For the reader's reference, we place two $10^{-10}$ normalized intensity ``Earths" at 60 miliarcseconds and 100 miliarcseconds on the focal plane, corresponding to roughly $3.5$ and $5.7 \lambda_0 / D$, respectively.
We also remind the reader that the results in Figure \ref{fig:550nm_focalplane_eac1} are subject to all polarization aberrations, there is no upstream polarizer that mitigates polarization aberration.
These data suggest that the nominal EAC-1 configuration was conservative in its design, and that a 12 meter barrel would in theory result in $10^{-10}$ normalized intensity at the inner working angle. We remind the reader of the disclaimer made in Section \ref{sec:methods}. These results are a \emph{lower bound} on the expected polarization-limited contrast floor, and do not consider other deleterious effects that degrade contrast at the IWA (e.g. stellar diameter, jitter). However, the method proposed in this work allows us to rapidly probe the solution space for scenarios that can be followed up with more rigorous physical optics modeling, including other deleterious effects.

\begin{figure}[ht]
    \centering
    \includegraphics[width=\linewidth, trim={0cm 0cm 0cm 0cm}]{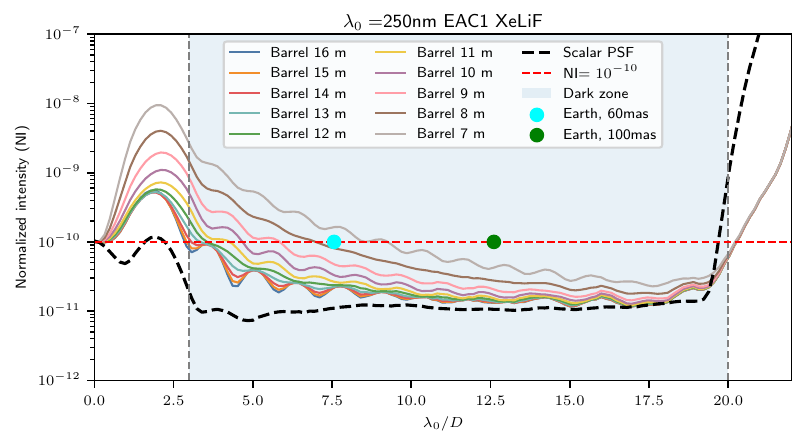}
    \caption{Azimuthally-averaged PSF profiles for each of the 10 optical designs used in this study, with polarization aberrations evaluated in the visible for a $10\%$ bandpass centered on 250nm. The profiles are plotted as a function of the center wavelength divided by the entrance pupil diameter, $\lambda_{0} / D$. To keep track of the chromatic scaling of the PSF, we also drop two fictitious ``Earths" at 60 and 100 mili-arcseconds (mas) to qualitatively assess how such exoplanets might be limited by polarization aberrations. In these data, we observe both Earths are far from the IWA of the coronagraph. Only the 8m barrel would in principle leave brighter residuals than the 60mas Earth, and the 100mas Earth is far enough from IWA that even the 7m barrel leaves polarized residuals.}
    \label{fig:250nm_focalplane_eac1}
\end{figure}

The results for visible wavelengths are encouraging, but they do not tell a complete story.
Polarization aberrations are generally larger in the ultraviolet, wherein lies the Ozone marker for characterizing the stage in an exoEarth's evolution. 
Pointing our analysis to the ultraviolet ($\lambda_0 = 250nm$, $\Delta \lambda = 10\%$) clarifies the impact of polarization aberrations at shorter wavelengths \cite{Will_polarization_luvoir, ashcraft_6mst_2025, ashcraft_gsmts_2025}.
Materials with normal dispersion tend to impart greater polarization aberrations at shorter wavelengths.
The greater polarization aberrations have been a concern for the designers of NASA's great observatories, suggesting that it will inhibit Earth-like exoplanet detection.
What we find in Figure \ref{fig:250nm_focalplane_eac1} suggests that this may not necessarily be the case. 
Yes, the polarization aberrations are greater than the case shown in Figure \ref{fig:550nm_focalplane_eac1}.
However, the linear scaling of the point-spread function works out in our favor for detecting the two fiducial planets.
The 60mas Earth is brighter than the focal plane residuals of every case but the 7m barrel, and the 100mas Earth is so far from the IWA that not even the 7 meter barrel produces polarization residuals that would obscure it.
An important caveat here is that there may be other factors (e.g. chromatic effects from surface roughness, scalar wavefront error) that limit contrast in the ultraviolet. However, this analysis suggests it is unlikely to be the polarization aberrations from thin-film coatings in isolation.

This is not the first work to suggest this effect.
Van Gorkom et al. 2025\cite{VanGorkom_2025_jatis} present a substantial error budgeting effort to estimate the performance limitations for a UV coronagraph onboard HWO, finding polarization to be a very modest error term.
Critical to consider in future work then is the contribution of form birefringence in the ultraviolet, as suggested by Breckinridge et al. 2018. Form birefringence is purely a function of the chemical composition and deposition of coatings, and can arise from the columnar stacking of metal nanoparticles. This effect may impart retardation that scales faster than the results shown in this study, so this effect must be considered\cite{Breckinridge_2018}.

Profiles for three other bands going toward the infrared are given in Appendix A for the reader's reference, but all tell the same story.
The polarization aberration residuals get smaller as the wavelength gets larger, and the ``Earths" get closer to the IWA as the wavelength gets larger.

\begin{figure}[ht]
    \centering
    \includegraphics[width=\linewidth]{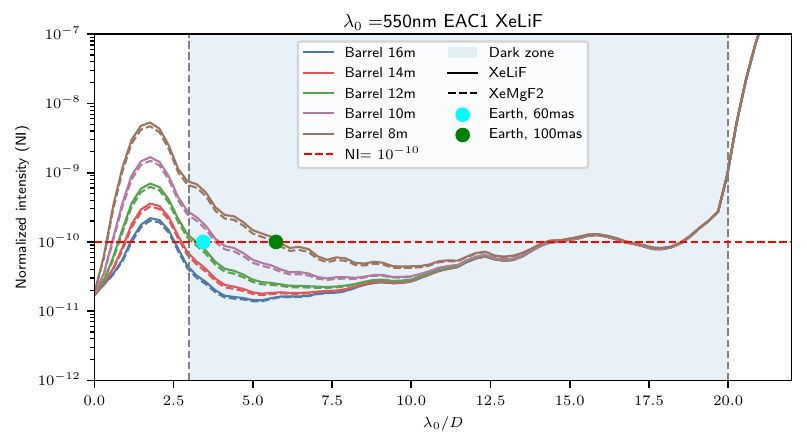}
    \caption{Azimuthally-averaged PSF profiles for each of the 10 optical designs used in this study, with polarization aberrations evaluated in the visible for a $20\%$ bandpass centered on 550nm, evaluated for both the XeLiF and XeMgF2 coatings. The profiles are plotted as a function of the center wavelength divided by the entrance pupil diameter, $\lambda_{0} / D$. Each color represents a different design, and each linestyle (i.e. solid, dashed) represents a different dielectric layer deposited on the primary and secondary mirrors. The difference in performance between the two coatings is marginal, with the XeMgF2 coating slightly outperforming the XeLiF coating.}
    \label{fig:550nm_focalplane_eac1_coating_compare}
\end{figure}

We don't observe a substantial change when simulating a different mirror coating on the primary and secondary. We repeat the analysis that produces Figure \ref{fig:550nm_focalplane_eac1}, switching the Al + XeLiF coating for an Al + XeMgF2 coating, which is shown in Figure \ref{fig:550nm_focalplane_eac1_coating_compare}. The two coatings don't substantially differ, with the XeMgF2 coating slightly outperforming the XeLiF coating. This isn't suprising, given their equivalent performance in Figure \ref{fig:coating_nominal}. At these wavelengths, the XeLiF coating has a slightly higher retardance, which directly correlates to larger contrast residuals \cite{ashcraft_6mst_2025}. We expect most aluminum-based coatings with a very thin dielectric top layer to behave similarly.

\begin{figure}[ht]
    \centering
    \includegraphics[width=\linewidth]{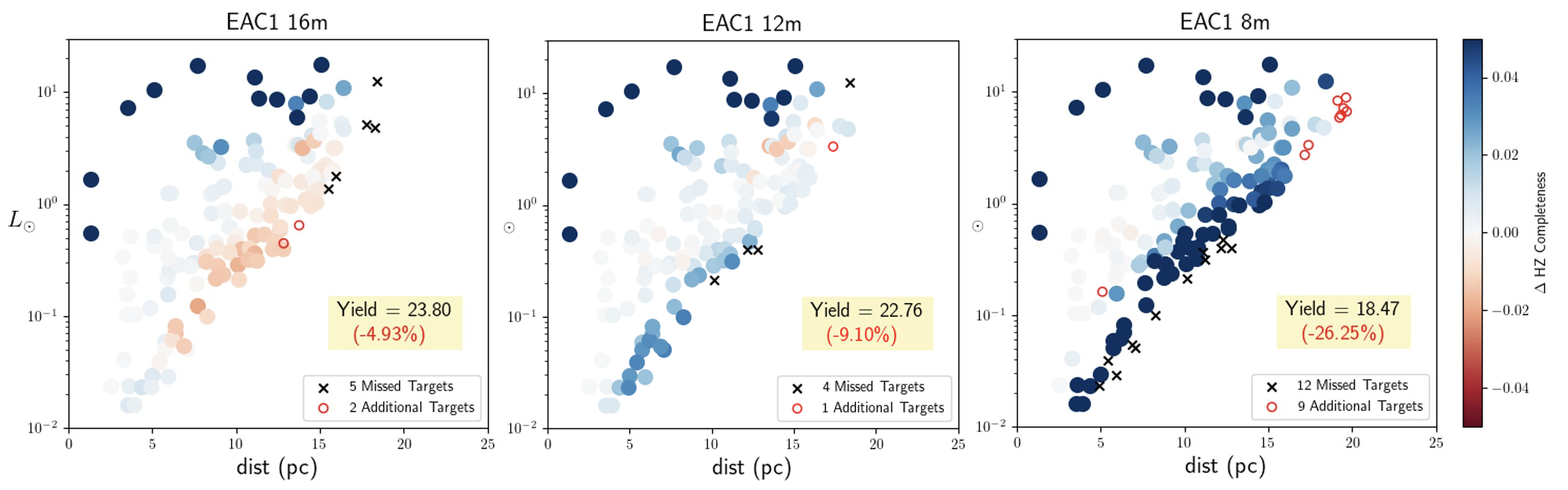}
    \caption{Yield diagrams showing the difference in HZ Completeness between a polarization aberrated case and the case without polarization aberration at 550nm. These data are plotted in such a way that a ``bluer" point has higher completeness in the polarization aberrated case, and a ``redder" point has higher completeness in the case without polarization aberration. The 16m case (left) shows that the marginal decrease in yield is being limited predominantly by the targets farther away (i.e., at a smaller angular separation). AYO is making up for the diminished completeness by ensuring a higher completeness on very luminous stars near the top of the plot. In the cases with higher polarization aberration (12m, center, 8m, right) we see that AYO starts to miss far-away targets around less luminous stars, and aims to drive the completeness higher to compensate for the decrease in yield. The colorbar on this plot is chosen to highlight the detail on the nominal EAC-1 design (16m).}
    \label{fig:compare_yield_scalar}
\end{figure}

The culmination of these analyses is to connect the physical optics simulations to Earth-like exoplanet yields, which is shown in Figure \ref{fig:compare_yield_scalar} for $\lambda=550nm$. The catalog used for this yield data is the Habitable Worlds Observatory Preliminary Input Catalog (HPIC\cite{Tuchow_2024}), which was the default catalogue for FRIDAY during this analysis. AYO defines a target distribution and then optimizes xposure time, number of re-visits, time between visits, and the specific stars selected for observation, to maximize the likelihood of detecting a planet in the systems' habitable zone. We generate a set of PSF realizations for every case shown in Figure \ref{fig:compare_yield_scalar}, and compare them against the contrast degradation as polarization aberrations increase. 

\begin{figure}[ht]
    \centering
    \includegraphics[width=\linewidth]{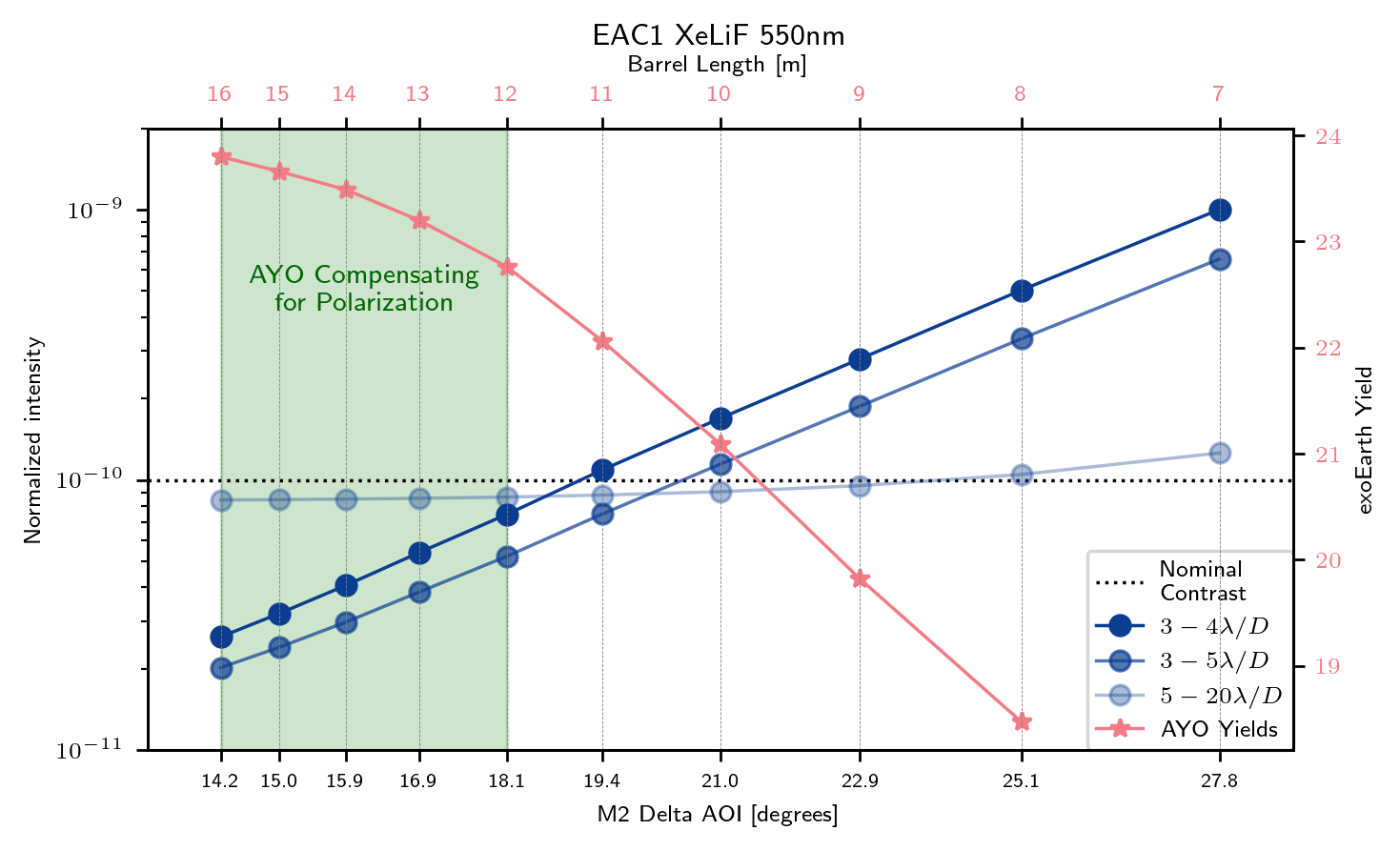}
    \caption{Plot of the degraded normalized intensity averaged over different annuli on the coronagraph focal plane as a function of the maximum change in angle of incidence ($\Delta\theta$) across the optical beam propagating through EAC-1. The left-most points correspond to the nominal EAC-1 design with polarization aberrations included. The blue plots show the degradation for $3-4\lambda/D$, $3-5\lambda/D$, and $5-20\lambda/D$. The first two, corresponding to low-spatial frequency errors, are the only plots that have a substantive change. Notably, it appears that the polarization aberrated normalized intensity near the IWA is linear in $\Delta\theta$. The same is not true for Earth-like exoplanet yield, which has a non-linear trend in the shaded region for the cases of smaller $\Delta\theta$. We interpret these results as AYO actively compensating for the small degradations imparted by polarization aberrations. This answers the question posed at the beginning of this study, there is a range where Yield Optimization can compensate for polarization aberration.}
    \label{fig:eac1_yield_contrast}
\end{figure}

The yields from FRIDAY tell an interesting story about how AYO maximizes HZ Completeness subject to varying amounts of polarization aberration. The coronagraph's IWA can be seen by the diagonal cutoff limiting targets to the upper left of the yield plots. This is where we see the greatest change in yield over the three simulated cases. The 16m case's modest polarization aberrations means that AYO can maximize yield by observing a few systems at large angular separations around relatively bright stars.
In the 12m case we observe AYO choosing to observe targets with a smaller angular separation around less-luminous stars to compensate for the degraded contrast at the IWA. The 8m case has a comparatively substantial degradation in yield that AYO is attempting to recover by observing targets near the coronagraph IWA with higher completeness. The small number of ``missed" or ``additional" in each plot arise from the shifting of the target distribution, and the fractional change in the distribution in minimal, so we don't consider this a significant effect. To finally answer the question posed in the Introduction, ``\emph{how sensitive is yield to polarization aberration?}", we evaluate the Yields for the remaining designs created for this study. We superimpose the Earth-like exoplanet yields over normalized intensity degradation in Figure \ref{fig:eac1_yield_contrast} to understand how these parameters trade with each other.

We observe two clear trends with these data. The first is that the degraded normalized intensity at the inner working angle is approximately linear as a function of the change in angle of incidence across the beam ($\Delta\theta$). As we expect, the average normalized intensity near the IWA increases with decreasing barrel length, or increasing $\Delta\theta$. The second trend is that exoEarth Yield is \emph{nonlinear} with $\Delta\theta$ for the cases with lower changes in angle of incidence. This clearly shows Altruistic Yield Optimization at work. AYO is compensating for the polarization aberrated normalized intensity by optimizing the design reference mission. This suggests that for small changes in $\Delta\theta$, yield is relatively insensitive to polarization aberration.

\section{RESULTS - EAC 4}
\label{sec:results-eac4}
As of the writing of this manuscript, the design process for the first three EAC's has concluded. Trades are ongoing for EAC-4 and 5, so it was of interest to the systems engineering team at the HWO project office to get an understanding for the polarization aberration limits set by the current EAC-4 design. All results henceforth should be treated as preliminary, given that the design of EAC-4 is not yet fixed. These results only include the influence of the EAC-4 OTA, as a ray trace model of the coronagraph is not yet available. EAC-4 is similarly a Ritchey-Chretien telescope design with a tertiary mirror that creates the coronagraph entrance pupil. In contrast to EAC-1, this observatory concept was designed with maximum angle of incidence (max AOI) in mind as the design specification. This resulted in 3 candidate OTA architectures, with gradually increasing max AOI to be conservative about polarization aberrations. The increasing max AOI was accomplished by increasing the off-axis distance of the mirror, which in turn results in a shorter barrel length. Our team was tasked with evaluating the evolution of polarization aberration with these three designs to determine if max AOI is a suitable specification to control polarization aberration.

\begin{figure}[ht]
    \centering
    \includegraphics[width=\linewidth]{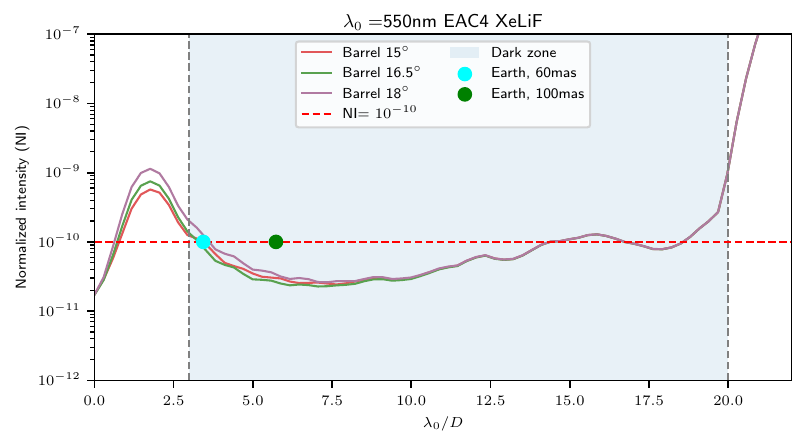}
    \caption{Azimuthally averaged focal plane residuals for EAC-4 for the three cases studied. Each design had a slightly different maximum angle of incidence, as reported in the Figure legend. We observe the trend that we expect, higher maximum angle of incidence yields higher polarization aberrations. In this case, because the change in angle of incidence is so small, the change in polarization aberration is marginal.}
    \label{fig:550nm_focalplane_eac4}
\end{figure}

We run the three designs through the same suite of analyses to determine the degradation of the focal plane subject to the varying angle of incidence for three specific cases. The results for this analysis are shown for the visible in Figure \ref{fig:550nm_focalplane_eac4}. We observe minimal change in the coronagraphic degradation between the three max AOI cases studied. Of note is that the nominal degradation with respect to the scalar case is higher than the results for the same wavelength and coating in EAC-1, shown in Figure \ref{fig:550nm_focalplane_eac1}. Table \ref{tab:compare_eac_params} includes a comparison of some different optical parameters to illustrate why this might be. The EAC-4 designs have a faster-focusing primary mirror, which result in a greater overall change in AOI across the beam. This directly correlates to increased polarization aberration. The EAC-4 perscriptions are similar in performance to the EAC-1 design at the 12m barrel made for this study. Indeed, the barrel lengths are of similar magnitude, and the focal ratios of the primary are similar. 


\begin{table}[H]
    \label{tab:compare_eac_params}
    \centering
    \begin{tabular}{c | c c c c c}
        \hline
        Specification & EAC-1 & EAC-1 (12m) & EAC-4A & EAC-4B & EAC-4C \\
        \hline
        Maximum AOI [deg] & 13.0 & 17.2 & 15.0 & 16.5 & 18 \\
        $\Delta$ AOI [deg] & 11.4 & 14.9 & 13.1 & 14.2 & 15.3 \\
        Primary Mirror Focal Ratio & F/2.47 & F/1.85 & F/2.18 & F/1.92 & F/1.76 \\
        Barrel Length [m] & 16 & 12 & $\approx$14 & $\approx$12 & $\approx$10 \\
        \hline
        
    \end{tabular}
    \caption{Relevant optical design specifications across EAC-1 and the three EAC-4 designs studied in this section. The approximate barrel length is included in the last row, where the EAC-4 Barrels have been rounded to the nearest EAC-1 barrel considered for comparison.}
\end{table}

\section{CONCLUSIONS}
\label{sec:conclusions}
The analysis presented in this work integrates a unique optical performance limiter with science yield via a high-fidelity wavefront sensing and control simulation. Using open-source tools for polarization ray tracing and jacobian-free wavefront control, we determine that the EAC-1's 16 meter barrel was conservative in the absence of polarization splitting. Subject to the assumptions in this study, polarization aberrations degrade the 12 meter barrel EAC-1 design (and shorter) in the visible ($\lambda=550nm$) to a normalized intensity of $>10^{-10}$ at the inner working angle. We remind the reader that this is the lower bound we can expect to achieve subject to a perfect coronagraph, and ideal wavefront sensing and control. We also find that, despite the ultraviolet containing a larger amplitude of polarization aberration, the small scale of the PSF with respect to expected angular separations of habitable exoplanets is well-suited to detecting exo-Earths. We study two different aluminum-based coatings and find that the differences in normalized intensity degradation are marginal. Given the simplicity of the XeLiF and XeMgF2 coatings, we expect similar coatings composed of aluminum with a thin dielectric layer to perform similarly. Connecting these analyses to AYO reveals that exoplanet yield calculators are capable of using astrophysical degrees of freedom, like exposure time and number of observations, to compensate for a range of polarization aberrations. Once again, after the 12 meter barrel exo-Earth yield begins to drop precipitously. We assess the degradation of normalized intensity of three point designs considered for EAC-4 with varying maximum angle of incidence. The residual normalized intensity degradation is similar across the three designs due to the modest change in angle of incidence. It is important for the reader to understand that the results presented here do not claim that polarization aberrations are the sole driving factor that will establish the barrel length of HWO. Other factors, such as appropriate stray light control (baffling) and micrometeoroid mitigation will also influence the final barrel length of HWO. One of the main outcomes of this work is a high-fidelity pipeline that determines the best contrast achievable for a given design subject to polarization aberrations.

Despite the high fidelity achieved in the constructed optical model, there were a few aspects of this analysis that merit future work.
First and foremost, the baseline coronagraph for EAC-1 is not yet physically realizable.
Notable work has been done to manufacture achromatic and scalar vorticies for HWO by Desai\cite{Desai_dimple_2024, Desai_wrapped_2023}, Palatnick\cite{Palatnick:24, Palatnick_lessons_learned_2024, Palatnick_manufacture_2024}, and K\"onig\cite{Konig_vortex_2025}\footnote{This list is non-exhaustive}, but a truly achromatic and scalar vortex mask is a significant technical challenge.
The ideal vortex model was suitable to assessing the contribution of polarization aberration to normalized intensity degradation. Our analysis can now be extended to including masks with dispersion and defects near the central singularity.
This pipeline is built to study other coronagraph architectures, like the Phase- and Amplitude-apodized Pupil Lyot Coronagraphs\cite{Por_2020, Riggs_2025}.
Low-order aberration sensitivity is known to vary with coronagraph architecture\cite{cds_pipeline}, so the results presented in this study will vary with coronagraph type. 

Other coronagraphs with smaller inner working angles may be subject to higher sensitivity to polarization aberration. As the architecture for HWO matures, so should polarization aberration metrology and compensation strategies. In 2024, a Mueller imaging polarimeter was used to characterize the polarization aberrations present in the SCoOB coronagraph\cite{ashcraft_scoob_2024}. However, these results were of questionable accuracy, and future work should refine the polarimetric calibration to ensure accurate results. Metasurface and liquid crystal polarization optics prove to be a promising technology for polarization aberration compensation. The ability to directly write a spatially-varying birefringent pattern onto an optic can nullify the retardance accrued by light as it propagates through HWO. 

To democratize the analysis in this work, we intend to release the polarization modeling tools as the open-source \verb|hwo_pol|. While this analysis tool is not yet a formally supported Python package, we invite members from the community to collaborate with us at the url \url{https://github.com/Jashcraf/hwo_pol}.

\section{Appendix A - Other EAC1 PSF Profiles}

\begin{figure}[H]
    \centering
    \includegraphics[width=\linewidth]{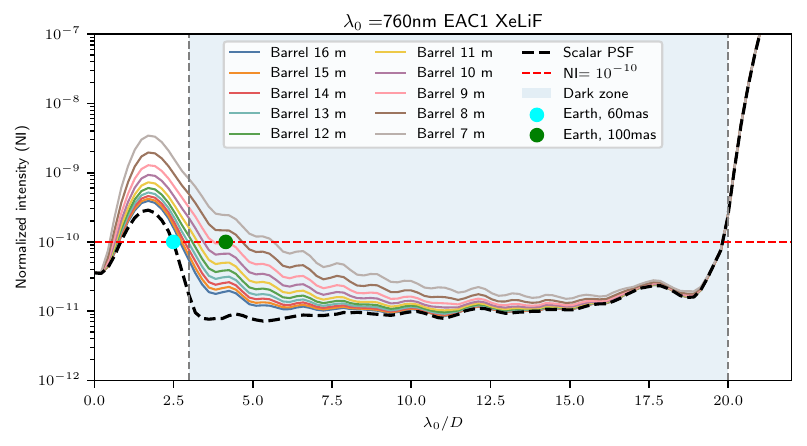}
    \caption{Azimuthally-averaged PSF profiles for each of the 10 optical designs used in this study, with polarization aberrations evaluated in the visible for a $20\%$ bandpass centered on 760nm. The profiles are plotted as a function of the center wavelength divided by the entrance pupil diameter, $\lambda_{0} / D$. To keep track of the chromatic scaling of the PSF, we also drop two fictitious ``Earths" at 60 and 100 mili-arcseconds (mas) to qualitatively assess how such exoplanets might be limited by polarization aberrations. In these data, the 60mas planet is beyond the IWA, but the 100mas planet is in principle visible to designs employing a 9m barrel or longer.}
\end{figure}

\begin{figure}[ht]
    \centering
    \includegraphics[width=\linewidth]{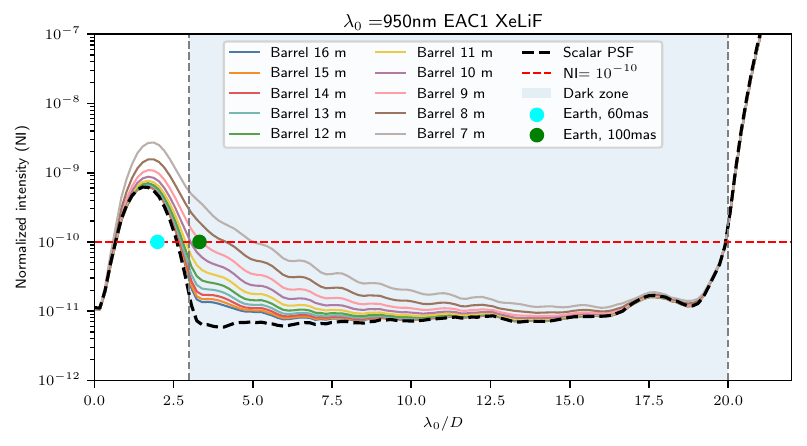}
    \caption{Azimuthally-averaged PSF profiles for each of the 10 optical designs used in this study, with polarization aberrations evaluated in the visible for a $20\%$ bandpass centered on 950nm. The profiles are plotted as a function of the center wavelength divided by the entrance pupil diameter, $\lambda_{0} / D$. To keep track of the chromatic scaling of the PSF, we also drop two fictitious ``Earths" at 60 and 100 mili-arcseconds (mas) to qualitatively assess how such exoplanets might be limited by polarization aberrations. In these data, the 60mas planet is beyond the IWA, and the 100mas planet is nearly at the IWA. However, it is at nearly the same contrast as the polarized degradation from the 9 meter case, indicating it would be visible for barrels longer than this.}
\end{figure}

\begin{figure}[ht]
    \centering
    \includegraphics[width=\linewidth]{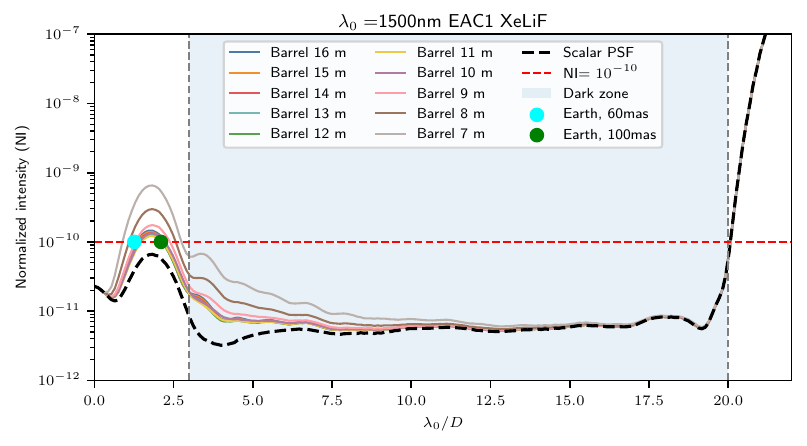}
    \caption{Azimuthally-averaged PSF profiles for each of the 10 optical designs used in this study, with polarization aberrations evaluated in the visible for a $20\%$ bandpass centered on 1500nm. The profiles are plotted as a function of the center wavelength divided by the entrance pupil diameter, $\lambda_{0} / D$. To keep track of the chromatic scaling of the PSF, we also drop two fictitious ``Earths" at 60 and 100 mili-arcseconds (mas) to qualitatively assess how such exoplanets might be limited by polarization aberrations. In these data, both Earths are beyond the coronagraph's inner working angle, making the EAC-1 coronagraph not suitable for this band. Consistent with previous literature, we find even lower polarization aberration here for even the most extreme case.}
\end{figure}

\section*{Code, Data, and Materials Availability Statement}
This research primarily made use of the following open-source software packages: \emph{Poke}\cite{Ashcraft_poke_2023}, \verb|hwo_sim|, and \verb|hwo_pol|. The \verb|hwo_pol| repository is hosted in Github \url{https://github.com/Jashcraf/hwo_pol}. Other data and analysis code can be provided upon reasonable request.

\section*{Disclosures}
The authors declare that there are no financial interests, commercial affiliations, or other potential conflicts of interest that could have influenced the objectivity of this research or the writing of this paper

\acknowledgments 

J.N.A was supported by NASA through the NASA Hubble Fellowship grant $\#$HST-HF2-51547.001-A awarded by the Space Telescope Science Institute, which is operated by the Association of Universities for Research in Astronomy. 
The authors would like to acknowledge other memebers of the Polarization PSDD for helpful discussions throughout this work, in no particular order: Rhonda Morgan, Brandon Dube, Pin Chen, John Trauger, Brian Fleming, Dimitri Savaransky, Emiel Por, and Ruslan Belikov.

\bibliographystyle{spiebib} 
\bibliography{report_new} 

\begin{thebibliography}{10}

\bibitem{Breckinridge_2015}
Breckinridge, J.~B., Lam, W. S.~T., and Chipman, R.~A., ``Polarization aberrations in astronomical telescopes: The point spread function,'' {\em Publications of the Astronomical Society of the Pacific}~{\bf 127},  445 (may 2015).

\bibitem{anche_2023}
{Anche, Ramya M.}, {Ashcraft, Jaren N.}, {Haffert, Sebastiaan Y.}, {Millar-Blanchaer, Maxwell A.}, {Douglas, Ewan S.}, {Snik, Frans}, {Williams, Grant}, {van Holstein, Rob G.}, {Doelman, David}, {Van Gorkom, Kyle}, and {Skidmore, Warren}, ``Polarization aberrations in next-generation giant segmented mirror telescopes (gsmts) - i. effect on the coronagraphic performance,'' {\em A\&A}~{\bf 672},  A121 (2023).

\bibitem{ashcraft_gsmts_2025}
{Ashcraft, Jaren N.}, {Anche, Ramya M.}, {Haffert, Sebastiaan Y.}, {Justin Hom}, {Millar-Blanchaer, Maxwell A.}, {Douglas, Ewan S.}, {Snik, Frans}, {van Holstein, Rob G.}, {Doelman, David}, {Van Gorkom, Kyle}, {Skidmore, Warren}, and {Manxuan Zhang}, ``Polarization aberrations in next-generation giant segmented mirror telescopes (gsmts) - ii. influence of segment-to-segment coating variations on high-contrast imaging and polarimetry,'' {\em A\&A}~{\bf 695},  A28 (2025).

\bibitem{Stark_14}
{Christopher C. Stark}, {Aki Roberge}, {Avi Mandell}, and {Tyler D. Robinson}, ``"maximizing the exoearth candidate yield from a future direct imaging mission",'' {\em ApJ}  (2014).

\bibitem{krist_numerical_2023}
Krist, J.~E., Steeves, J.~B., Dube, B.~D., Riggs, A.~E., Kern, B.~D., Marx, D.~S., Cady, E.~J., Zhou, H., Poberezhskiy, I.~Y., Baker, C.~W., McGuire, J.~P., Nemati, B., Kuan, G.~M., Mennesson, B., Trauger, J.~T., Saini, N.~S., and Rafels, S.~H., ``{End-to-end numerical modeling of the Roman Space Telescope coronagraph},'' {\em Journal of Astronomical Telescopes, Instruments, and Systems}~{\bf 9}(4),  045002 (2023).

\bibitem{ashcraft_6mst_2025}
Ashcraft, J.~N., Dube, B.~D., Douglas, E.~S., Kim, D., Krist, J.~E., Mennesson, B., Monacelli, B., Morgan, R., Raouf, N.~A., Riggs, A.~E., Rodgers, M., and Warfield, K.~R., ``{Comparison of polarization aberrations from existing mirror coatings for coronagraphic imaging of habitable worlds},'' {\em Journal of Astronomical Telescopes, Instruments, and Systems}~{\bf 11}(1),  015002 (2025).

\bibitem{Feinberg2024}
Feinberg, L., Ziemer, J., Ansdell, M., Crooke, J., Dressing, C., Mennesson, B., O'Meara, J., Pepper, J., and Roberge, A., ``{The Habitable Worlds Observatory engineering view: status, plans, and opportunities},'' in [{\em Space Telescopes and Instrumentation 2024: Optical, Infrared, and Millimeter Wave}{\nolinebreak\hspace{0.1em}]},  Coyle, L.~E., Matsuura, S., and Perrin, M.~D., eds.,  {\bf 13092},  130921N, International Society for Optics and Photonics, SPIE (2024).

\bibitem{Ashcraft_poke_2023}
Ashcraft, J.~N., Douglas, E.~S., Kim, D., Riggs, A. J.~E., Anche, R., Brendel, T., Derby, K., Dube, B.~D., Jarecki, Q., Jenkins, E., and Milani, K.~S., ``{Poke: an open-source, ray-based physical optics platform},'' in [{\em Optical Modeling and Performance Predictions XIII}{\nolinebreak\hspace{0.1em}]},  Kahan, M.~A., ed.,  {\bf 12664},  1266404, International Society for Optics and Photonics, SPIE (2023).

\bibitem{CLY}
Chipman, R.~A., Lam, W.-S.~T., and Young, G.,  [{\em Polarized Light and Optical Systems}{\nolinebreak\hspace{0.1em}]}, CRC Press (2018).

\bibitem{Yun:11_1}
Yun, G., Crabtree, K., and Chipman, R.~A., ``Three-dimensional polarization ray-tracing calculus i: definition and diattenuation,'' {\em Appl. Opt.}~{\bf 50},  2855--2865 (Jun 2011).

\bibitem{Yun:11_2}
Yun, G., McClain, S.~C., and Chipman, R.~A., ``Three-dimensional polarization ray-tracing calculus ii: retardance,'' {\em Appl. Opt.}~{\bf 50},  2866--2874 (Jun 2011).

\bibitem{Will_polarization_luvoir}
Will, S.~D. and Fienup, J.~R., ``{Effects and mitigation of polarization aberrations in LUVOIR coronagraph},'' in [{\em Techniques and Instrumentation for Detection of Exoplanets IX}{\nolinebreak\hspace{0.1em}]},  Shaklan, S.~B., ed.,  {\bf 11117},  1111710, International Society for Optics and Photonics, SPIE (2019).

\bibitem{Pueyo:09}
Pueyo, L., Kay, J., Kasdin, N.~J., Groff, T., McElwain, M., Give'on, A., and Belikov, R., ``Optimal dark hole generation via two deformable mirrors with stroke minimization,'' {\em Appl. Opt.}~{\bf 48},  6296--6312 (Nov 2009).

\bibitem{Will_jacobian_2021}
Will, S.~D., Groff, T.~D., and Fienup, J.~R., ``{Jacobian-free coronagraphic wavefront control using nonlinear optimization},'' {\em Journal of Astronomical Telescopes, Instruments, and Systems}~{\bf 7}(1),  019002 (2021).

\bibitem{exosims}
{Savransky}, D., {Delacroix}, C., and {Garrett}, D., ``{EXOSIMS: Exoplanet Open-Source Imaging Mission Simulator}.'' Astrophysics Source Code Library, record ascl:1706.010 (June 2017).

\bibitem{cds_pipeline}
Belikov, R., Stark, C., Siegler, N., Por, E., Mennesson, B., Redmond, S., Chen, P., Fogarty, K., Guyon, O., Juanola-Parramon, R., Kasdin, J., Krist, J., Mawet, D., Morgan, R., Prada, C.~M., Pueyo, L., Ruane, G., Sirbu, D., Stapelfeldt, K., Trauger, J., Zimmerman, N., Alagao, M. A.~M., Carlotti, A., Chafi, J., Doleman, D., Gersh-Range, J., K{\"o}nig, L., Leboulleux, L., Moody, D., Riggs, A.~J., Serabyn, E., Snik, F., and Wallace, K., ``{Coronagraph design survey for future exoplanet direct imaging space missions},'' in [{\em Space Telescopes and Instrumentation 2024: Optical, Infrared, and Millimeter Wave}{\nolinebreak\hspace{0.1em}]},  Coyle, L.~E., Matsuura, S., and Perrin, M.~D., eds.,  {\bf 13092},  1309266, International Society for Optics and Photonics, SPIE (2024).

\bibitem{Will_2025_idealized}
Will, S.~D., Ashcraft, J.~N., Stark, C.~C., Millar-Blanchaer, M.~A., and Sitarski, B.~N., ``Fundamental limits to coronagraphic wavefront correction with pairwise-type estimation in the presence of polarization aberrations,'' {\em Opt. Express}~{\bf 33},  51871--51886 (Dec 2025).

\bibitem{feinberg2026habitableworldsobservatorysconcept}
Feinberg, L.~D., Sitarski, B.~N., McElwain, M.~W., Arney, G., Baker, C., Bolcar, M.~R., Levine, M., Liu, A., Mennesson, B., Roberge, A., Smith, J.~S., Zhao, F., and Ziemer, J., ``Habitable worlds observatory's concept and technology maturation: Initial feasibility and trade space exploration,'' (2026).

\bibitem{Doelman_2022}
Doelman, D.~S., Ouellet, M., Ruane, G., Escuti, M., Haffert, S., and Snik, F., ``{First laboratory tests of a triple-grating vector vortex coronagraph},'' in [{\em Space Telescopes and Instrumentation 2022: Optical, Infrared, and Millimeter Wave}{\nolinebreak\hspace{0.1em}]},  Coyle, L.~E., Matsuura, S., and Perrin, M.~D., eds.,  {\bf 12180},  1218029, International Society for Optics and Photonics, SPIE (2022).

\bibitem{Palatnick:24}
Palatnick, S., Millar-Blanchaer, M.~A., Wallace, J.~K., John, D.~D., Moore, A., and Wenger, T., ``Achromatizing photolithographically patterned metasurfaces with arbitrary, variable unit cell size,'' {\em Opt. Express}~{\bf 32},  47057--47075 (Dec 2024).

\bibitem{Quijada_2025}
Quijada, M.~A., Hoyo, J. G.~D., Marcos, L. V. R.~D., Wollack, E.~J., Batkis, M.~F., Lewis, D.~M., Rydalch, T.~D., and Allred, D.~D., ``{Far-ultraviolet optical properties and performance of physical vapor deposited aluminum mirrors protected with XeLiF and XeMgF2},'' {\em Journal of Astronomical Telescopes, Instruments, and Systems}~{\bf 11}(4),  042209 (2025).

\bibitem{Noecker_hwo}
Noeker, C. et~al., ``Coronagraph instrument for habitable worlds observatory,'' {\em this volume}  (in press).

\bibitem{VanGorkom_2025_jatis}
Gorkom, K. J.~V., Anche, R.~M., Mendillo, C.~B., Gersh-Range, J.~A., Hom, J., Robinson, T.~D., N'Diaye, M., Lewis, N.~K., Macintosh, B.~A., and Douglas, E.~S., ``{Performance predictions and contrast limits for an ultraviolet high-contrast imaging testbed},'' {\em Journal of Astronomical Telescopes, Instruments, and Systems}~{\bf 11}(4),  042203 (2025).

\bibitem{Breckinridge_2018}
Breckinridge, J.~B., Kupinski, M., Davis, J., Daugherty, B., and Chipman, R.~A., ``{Terrestrial exoplanet coronagraph image quality polarization aberrations in Habex},'' in [{\em Space Telescopes and Instrumentation 2018: Optical, Infrared, and Millimeter Wave}{\nolinebreak\hspace{0.1em}]},  Lystrup, M., MacEwen, H.~A., Fazio, G.~G., Batalha, N., Siegler, N., and Tong, E.~C., eds.,  {\bf 10698},  106981D, International Society for Optics and Photonics, SPIE (2018).

\bibitem{Tuchow_2024}
Tuchow, N.~W., Stark, C.~C., and Mamajek, E., ``Hpic: The habitable worlds observatory preliminary input catalog,'' {\em The Astronomical Journal}~{\bf 167},  139 (feb 2024).

\bibitem{Desai_dimple_2024}
Desai, N., Mawet, D., Serabyn, E., Ruane, G., Bertrou-Cantou, A., Llop-Sayson, J., and Riggs, A. J.~E., ``{Benefits of adding radial phase dimples on scalar coronagraph phase masks},'' {\em Journal of Astronomical Telescopes, Instruments, and Systems}~{\bf 10}(1),  015001 (2024).

\bibitem{Desai_wrapped_2023}
Desai, N., Ruane, G.~J., Llop-Sayson, J.~D., Betrou-Cantou, A., Potier, A., Riggs, A.~E., Serabyn, E., and Mawet, D.~P., ``{Laboratory demonstration of the wrapped staircase scalar vortex coronagraph},'' {\em Journal of Astronomical Telescopes, Instruments, and Systems}~{\bf 9}(2),  025001 (2023).

\bibitem{Palatnick_lessons_learned_2024}
Palatnick, S., K{\"o}nig, L., Millar-Blanchaer, M., Wallace, J.~K., Serabyn, E., Mawet, D., Desai, N., John, D., and Schuller, J.~A., ``{Optimizing metasurfaces to achieve deeper direct imaging contrasts: analyses of current performance and lessons learned from fabrication},'' in [{\em Advances in Optical and Mechanical Technologies for Telescopes and Instrumentation VI}{\nolinebreak\hspace{0.1em}]},  Navarro, R. and Jedamzik, R., eds.,  {\bf 13100},  1310063, International Society for Optics and Photonics, SPIE (2024).

\bibitem{Palatnick_manufacture_2024}
Palatnick, S., John, D., and Millar-Blanchaer, M., ``Investigating pathways for deep-uv photolithography of large-area nanopost-based metasurfaces with high feature-size contrast,'' {\em Journal of Vacuum Science \& Technology B}~{\bf 42},  062602 (10 2024).

\bibitem{Konig_vortex_2025}
K{\"o}nig, L., Desai, N., Palatnick, S., Absil, O., Mawet, D., Millar-Blanchaer, M., and Serabyn, E., ``{Microstructured vortex and azimuthal cosine phase mask design for high-contrast imaging},'' {\em Journal of Astronomical Telescopes, Instruments, and Systems}~{\bf 11}(2),  025002 (2025).

\bibitem{Por_2020}
Por, E.~H., ``Phase-apodized-pupil lyot coronagraphs for arbitrary telescope pupils,'' {\em The Astrophysical Journal}~{\bf 888},  127 (jan 2020).

\bibitem{Riggs_2025}
Riggs, A. J.~E., Bailey, V.~P., Moody, D., Balasubramanian, K., Basinger, S.~A., Belikov, R., Bendek, E., Debes, J., Dube, B.~D., Gersh-Range, J., Groff, T.~D., Kasdin, N.~J., Mennesson, B., Monacelli, B., Moore, D.~M., Ruane, G., Sandhu, J., Shi, F., Sidick, E., Siegler, N., Sirbu, D., Trauger, J., Weisberg, C.~L., White, V.~E., Wilson, D.~W., Wilson, R.~C., Yee, K.~Y., and Zimmerman, N.~T., ``{Flight masks of the Roman Space Telescope Coronagraph Instrument},'' {\em Journal of Astronomical Telescopes, Instruments, and Systems}~{\bf 11}(2),  021403 (2025).

\bibitem{ashcraft_scoob_2024}
Ashcraft, J.~N., Douglas, E.~S., Anche, R.~M., Gorkom, K.~V., Jenkins, E., Melby, W., and Millar-Blanchaer, M.~A., ``{The space coronagraph optical bench (SCoOB): 3. Mueller matrix polarimetry of a coronagraphic exit pupil},'' in [{\em Space Telescopes and Instrumentation 2024: Optical, Infrared, and Millimeter Wave}{\nolinebreak\hspace{0.1em}]},  Coyle, L.~E., Matsuura, S., and Perrin, M.~D., eds.,  {\bf 13092},  130926K, International Society for Optics and Photonics, SPIE (2024).

\end{thebibliography}

\end{document}